\documentclass[conference]{IEEEtran}

\IEEEoverridecommandlockouts
\usepackage{cite}
\usepackage{amsmath,amssymb,amsfonts}
\usepackage{algorithm,algorithmic}
\usepackage{graphicx}
\usepackage{textcomp}
\usepackage{xcolor}
\usepackage{hyperref}
\usepackage{url}
\usepackage{geometry}
\usepackage[caption=false,font=footnotesize]{subfig}

\def\BibTeX{{\rm B\kern-.05em{\sc i\kern-.025em b}\kern-.08em
    T\kern-.1667em\lower.7ex\hbox{E}\kern-.125emX}}

\begin{document}

\title{ISAC-Enabled Multi-UAV Collaborative Target Sensing for Low-Altitude Economy
}

\author{\IEEEauthorblockN{Rui Wang\textsuperscript{1}, Kaitao Meng\textsuperscript{2}, Deshi Li\textsuperscript{1}, and Liang Xu\textsuperscript{1}}
\IEEEauthorblockA{\textsuperscript{1}Electronic Information School, Wuhan University, Wuhan, China.\\
\textsuperscript{2}Department of Electrical and Electronic Engineering, The University of Manchester, Manchester, UK.\\
Emails: \textsuperscript{1}\{ruiwang, dsli, lgxu\}@whu.edu.cn, \textsuperscript{2}kaitao.meng@manchester.ac.uk}
}

\maketitle

\begin{abstract}
Integrated sensing and communication (ISAC) has attracted growing research interests to facilitate the large-scale development of the low-altitude economy (LAE).
However, the high dynamics of low-altitude targets may overwhelm fixed ISAC systems, particularly at the edge of their coverage or in blind zones.
Driven by high flexibility, unmanned aerial vehicle (UAV)-assisted ISAC can provide more freedom of design to enhance communication and sensing abilities. 
In this paper, we propose an ISAC-enabled multi-UAV dynamic collaborative target sensing scheme, 
where UAVs can dynamically adjust their flight and resource allocation for cooperative sensing of mobile target through communicating with the terrestrial cellular network with ISAC signals.
To achieve the precise sensing of the dynamic target, the posterior Cram\'{e}r-Rao bound (PCRB) for the target state is derived.
Subsequently, the PCRB minimization problem is formulated by jointly optimizing the UAV-BS association, UAVs' trajectories and bandwidth allocation, subject to the communication requirements for the UAVs.
However, the problem is challenging since it involves non-convex and implicit objective function with coupled optimization variables.
For a fast implementation of sensing and tracking, we propose a low-complexity iterative algorithm that can efficiently obtain a sub-optimal solution to the problem.
Specifically, the UAV-BS association is first determined by the communication-optimal solution.
Then the UAVs’ trajectories and bandwidth allocation are alternatively optimized based on the descent direction search algorithm.
Finally, numerical results are provided to validate the superiority of our proposed designs as compared to various benchmarks.

\end{abstract}


\section{Introduction}
\IEEEPARstart{T}{he} 
integration of unmanned aerial vehicles (UAVs) and 6G facilitates the emergence of low-altitude economy (LAE), which promotes a series of low-altitude applications, including logistics transportation, urban management, energy inspection, etc \cite{Jiang2025Integrated}.
To support the development of LAE, 6G network is expected to achieve enhanced communication ability for aerial users and target sensing ability for safety guarantee \cite{meng2025ISAC}.
Specifically, for authorized UAVs, typically with high-speed data transmission requirement \cite{Wang2025multi}, it is necessary to ensure high-quality wireless links between them and the terrestrial cellular network. 
In contrast, for unauthorized UAVs or illegal targets, since they may interfere and even threaten urban safety \cite{Al2023Detection}, real-time and precise sensing of such targets is also crucial.
Therefore, it is essential to provide high-quality communication and sensing services for the large-scale and safe development of LAE.

Driven by the advantages of spectrum and hardware sharing, integrated sensing and communication (ISAC), as a unified framework providing wireless communication and radar sensing functions, has been one of the key technologies in 6G networks \cite{liu2022ISAC}.
In the literature, there have been various works that investigate the exploitation of ISAC to enhance communication and sensing abilities \cite{pang2024dynamic}.
For example, in \cite{Li2024Towards}, the authors studied the region sensing coverage in a multi-static ISAC system, where the beamforming is optimized to maximize the detection probability of the prescribed region subject to the signal-to-interference-plus-noise ratio (SINR) requirement for each user.
Different from the fixed deployment of ground BSs, UAV-assisted ISAC systems provide a new degree of freedom (DoF) due to their flexibility \cite{meng2024UAV}. 
In \cite{Ding2023Multi}, the authors aimed at maximizing the minimum detection probability over a target area by jointly optimizing the locations and transmission power of UAVs while guaranteeing the communication requirements of users.
However, the above works mainly focus on providing sensing services for given locations or area.
In practice, due to the position's uncertainty and mobility of the illegal target, it is still challenging for the ISAC system to achieve continuous sensing of the dynamic target.

To satisfy the communication and sensing requirements in highly dynamic low-altitude environment, there are several challenges that need to be addressed.
Firstly, the target's location is unknown and dynamic, which makes it impossible to pre-design the trajectory and resource allocation of UAV. Instead, it needs to be updated in real time according to the current state of targets.
Secondly, the target sensing performance is dictated by the close-coupled UAVs' trajectories and bandwidth resources. A more orthogonal geometric relationship between the UAVs and the target contributes to enhanced sensing performance. However, the fixed distribution of BSs cannot guarantee a high transmission rate for the UAV at all locations.
Therefore, it is highly complex to obtain the optimal cooperative trajectories of UAVs and bandwidth allocation.

With the above consideration, we propose an ISAC-enabled multi-UAV dynamic collaborative target sensing scheme, where multiple authorized UAVs exploit ISAC signals to conduct sensing of the mobile target while communicating with the terrestrial BSs. 
To achieve the precise sensing of the dynamic target, the posterior Cram\'{e}r-Rao bound (PCRB) for the target state is derived as the sensing error metric. 
Then a PCRB minimization problem is formulated by jointly optimizing the UAV-BS association, UAVs' trajectories, and bandwidth allocation subject to the communication requirements for the UAVs.
Since the problem involves non-convex and implicit objective function with coupled optimization variables, we propose a low-complexity iterative algorithm that can efficiently obtain a sub-optimal solution.
First, a communication-optimal solution is obtained to check the feasibility and determine the UAV-BS association.
Then, the alternating optimization (AO) framework is employed to optimize the UAV trajectory and bandwidth allocation based on the descent direction search algorithm.
The main contributions are summarized as follows:
\begin{itemize}
	\item We propose an ISAC-enabled multi-UAV dynamic collaborative target sensing scheme, where UAVs can dynamically adjust their flight and resource allocation for cooperative sensing of mobile target. 
	\item We develop a low-complexity iterative algorithm for joint optimization of the UAV-BS association, UAVs’ trajectories and bandwidth allocation to minimize the PCRB of the target state, while satisfying the transmission requirement and reliable control of the UAVs.
	\item Simulation results show that the proposed scheme can effectively improve the sensing performance and reduce the estimated error in target location and velocity.
\end{itemize}


\section{System Model and Problem Formulation}
\begin{figure}[t]
	\centering{\includegraphics[width=7.2cm]{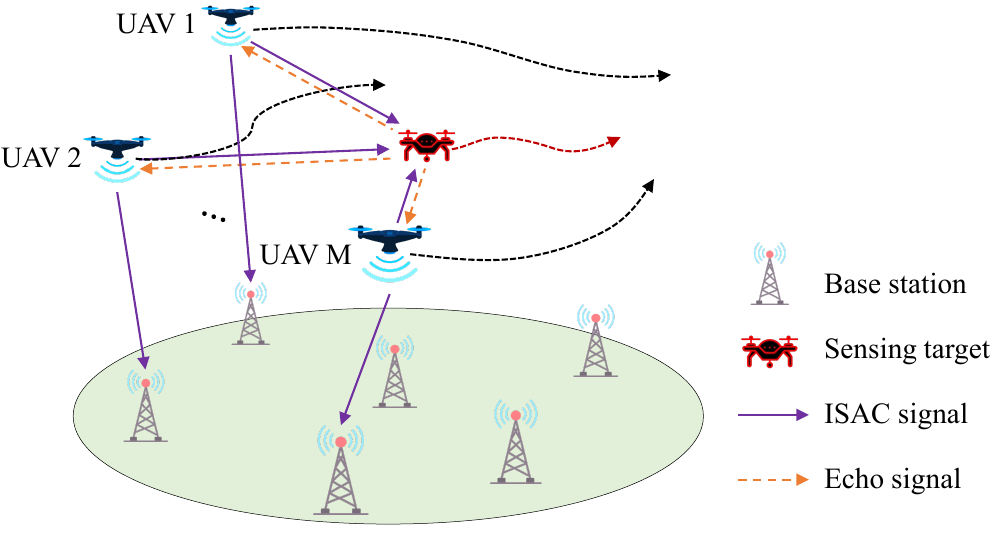}}
	\vspace{-3.5mm}
	\caption{ISAC-enabled multi-UAV collaborative sensing.}
    \vspace{-0.2cm}
	\label{system_model}
\end{figure}

As shown in Fig. \ref{system_model}, consider an ISAC-enabled multi-UAV collaborative sensing scenario consisting of $M$ UAVs, $K$ BSs, and one target.
The UAVs transmit the obtained sensing data to BSs, while reusing the information signals for target sensing at the region of interest.
The BSs not only receive the sensing data from UAVs for further processing, but also manage the wireless resources and flight of UAVs.
The target might be an unauthorized UAV or an illegal intruder, whose state such as location and velocity is unknown and dynamic.
The sets of UAVs and BSs are denoted by $\mathcal{M}=\{1,...,M\}$ and $\mathcal{K}=\{1,...,K\}$, respectively.
Assume that the UAVs and target fly at a constant altitude $H$.
The $k$-th BS's horizontal location is denoted by $\mathbf{w}_k \in \mathbb{R}^{2 \times 1}$, and the height of BS is $H^\mathrm{B}$.

\subsection{Communication Model}
Denote the duration of the ISAC service by $T$ and discretize it into $N$ tiny time intervals each with a length of $\Delta t=T/N$.
Thus the horizontal position of the $m$-th UAV at time $n$ is denoted as $\mathbf{q}_{m,n}=(x^u_{m,n},y^u_{m,n})$.
The UAV-BS communication link is assumed to be dominated by the line-of-sight (LoS) component.
Therefore, the channel power gain between the UAV $m$ and the BS $k$ at the time $n$ can be modeled as 
\begin{equation}
	h_{m,k,n} = \frac{\beta_0}{d_{m,k,n}^2} = \frac{\beta_0}{\|\mathbf{q}_{m,n}-\mathbf{w}_k\|^2 + \Delta H^2},
\end{equation}
where $\beta_0$ is the reference channel power gain at 1m distance, and $\Delta H = H-H^\mathrm{B}$.

To avoid interference among the UAVs, the frequency division multiple access (FDMA) is adopted. 
Denote the bandwidth allocation ratio of UAV $m$ at time interval $n$ as $\eta_{m,n}$, which satisfy $\sum_{m=1}^M \eta_{m,n}=1$ and $0\leq \eta_{m,n} \leq1, m\in\cal{M}$.
Denote the UAV-BS association by $a_{m,k,n}\in \{0,1\}$, where $a_{m,k,n}=1$ indicates that the $m$-th UAV is associated with the $k$-th BS at the $n$-th time interval, and $a_{m,k,n}=0$ otherwise.
Without loss of generality, We assume that each UAV can only communicate with one BS within each time interval, and each BS can serve $N_a$ UAVs.
Therefore, the achievable transmission rate of the UAV $m$ at the time interval $n$ can be given by
\begin{equation}
	R_{m,n}=\sum_{k=1}^K a_{m,k,n}\eta_{m,n}B\log_2\left(1+\frac{P^\text{u}_t h_{m,k,n}}{\sigma_0^2}\right),
\end{equation}
where $B$ is the total available bandwidth, $P^\text{u}_t$ is the transmission power of UAVs, and $\sigma_0^2$ is the power of channel noise.

Furthermore, to ensure the real-time and reliable control of the UAVs, the quality of the BS-UAV link is specified by a minimum received SNR requirement \cite{Zhang2019Cellular}.
The SNR received by the UAV $m$ from the BS $k$ at the time interval $n$ can be expressed as $\Gamma_{m,k,n}=\frac{P^\mathrm{BS}_t h_{m,k,n}}{\sigma_0^2}$, where $P^\mathrm{BS}_t$ is the transmission power of BSs.
Thus, it should satisfy 
$\min_k \Gamma_{m,k,n} \geq \Gamma_\mathrm{th}$, 
where $\Gamma_\mathrm{th}$ is the received SNR threshold.

\subsection{Sensing Model}
\subsubsection{Target State Evolution Model}
The target state at time interval $n$ is represented as $\mathbf{x}_n = [x^t_n, y^t_n, \dot{x}^t_n, \dot{y}^t_n]^T$, where $[x^t_n, y^t_n]^T$ and $[\dot{x}^t_n, \dot{y}^t_n]^T$ denote the location and velocity of the target respectively.
For simplicity, assuming that the target moves with a nearly constant velocity \cite{na2019joint}, then the target state evolution model can be summarized as 
\begin{equation}
	\mathbf{x}_n = \mathbf{F}\mathbf{x}_{n-1} + \boldsymbol{\omega}_{n-1},
\end{equation}
where 
$\mathbf{F} = 
\left[
\begin{array}{cc}
	1 & \Delta t \\
	0 & 1
\end{array}
\right]
\otimes \mathbf{I}_2,$
$\Delta t$ is the observation time interval, $\mathbf{I}_2$ denotes the $2\times 2$ identity matrix. 
$\boldsymbol{\omega}_{n-1} \sim \mathcal{N}(0, \mathbf{W})$ is the process noise, which is modeled as the Gaussian distribution with zero-mean and covariance matrix $\mathbf{W}$:
\begin{equation}
	\mathbf{W} = \kappa
	\left[
	\begin{array}{cc}
		\frac{1}{3}\Delta t^3 & \frac{1}{2}\Delta t^2 \\
		\frac{1}{2}\Delta t^2 & \Delta t
	\end{array}
	\right]
	\otimes \mathbf{I}_2,
\end{equation}
where $\kappa$ is the intensity of the process noise.

\subsubsection{Measurement Model}
The UAVs can obtain the range and velocity measurements of the target from the filtered echo signals.
Denote $\mathbf{y}_n=[d_{1,n},...,d_{M,n},v_{1,n},...,v_{M,n}]^T$ as the measurement vector at the time interval $n$.
The measurement model is described as 
\begin{equation}
	\mathbf{y}_n = \mathbf{h}(\mathbf{x}_n)+\mathbf{z}_n,
\end{equation}
where $\mathbf{h}(\cdot)$ is defined by
\begin{equation}
	\left\{
	\begin{aligned}
		& d_{m,n}=\sqrt{(x^t_n-x^u_{m,n})^2+(y^t_n-y^u_{m,n})^2}, \\
		& v_{m,n}=\frac{(x^t_n-x^u_{m,n})\dot{x}^t_n+(y^t_n-y^u_{m,n})\dot{y}^t_n}{d_{m,n}}.
	\end{aligned}
	\right.
\end{equation}
And $\mathbf{z}_n \sim \mathcal{N}(0,\mathbf{R}_n)$ is the measurement noise with the covariance matrix being expressed as
\begin{equation}
	\mathbf{R}_n = \mathrm{diag}(\sigma^2_{d_{1,n}},...,\sigma^2_{d_{M,n}},\sigma^2_{v_{1,n}},...,\sigma^2_{v_{M,n}}).
\end{equation}
The variances of the measurement noises are related to the SNR in the echo signal. The echo signal SNR in the UAV $m$ at the time interval $n$ can be obtained as
$\mathrm{SNR}_{m,n}^s = \frac{\lambda^2 P^\text{u}_t \sigma_\mathrm{RCS} }{(4\pi)^3 d_{m,n}^4 \sigma_0^2}$,
where $\sigma_\mathrm{RCS}$ is the radar cross section (RCS) of the target, and $\lambda$ is the carrier wavelength.
Thus the variances of the measurement noises satisfy \cite{zhang2020power}
\begin{equation}
	\left\{
	\begin{aligned}
		& \sigma^2_{d_{m,n}} = \frac{\alpha_d}{\mathrm{SNR}_{m,n}^s(\eta_{m,n}B)^2}, \\
		& \sigma^2_{v_{m,n}} = \frac{\alpha_v}{\mathrm{SNR}_{m,n}^s(T^s_{m,n})^2},
	\end{aligned}
	\right.
	\label{measurement noise}
\end{equation}
where $\alpha_d$ and $\alpha_v$ respectively represent the proportional coefficients of distance measurement error and velocity measurement error, $T^s_{m,n}$ is the effective measurement time duration.

\subsubsection{Sensing Performance Metric}
To evaluate the sensing performance of the target, the PCRB is derived, which provides a lower bound for the variances of any unbiased estimators \cite{liu2022radar}.
If let $\hat{\mathbf{x}}_n$ be an unbiased estimate of the target state using the measurement vector $\mathbf{y}_n$, the corresponding PCRB can be written as 
\cite{tichavsky1998posterior}
\begin{equation}
	\mathbb{E}_{\mathbf{x}_n,\mathbf{y}_n} \left[\left(\hat{\mathbf{x}}_n(\mathbf{y}_n)-\mathbf{x}_n\right)\left(\hat{\mathbf{x}}_n(\mathbf{y}_n)-\mathbf{x}_n\right)^T\right] \geq \mathbf{J}^{-1} \triangleq \mathbf{C} ,
\end{equation}
where $\mathbf{C}$ is the PCRB matrix, and $\mathbf{J}$ is the posterior Fisher information matrix (FIM), which can be calculated by
\begin{equation}
	\mathbf{J} 
	\!=\! -\mathbb{E}\left(\!\frac{\partial^2\ln{p(\mathbf{x}_n)}}{\partial \mathbf{x}_n^2}\!\right)
	\!-\!\mathbb{E}\left(\!\frac{\partial^2\ln{p(\mathbf{y}_n|\mathbf{x}_n)}}{\partial \mathbf{x}_n^2}\!\right)
	\!=\! \mathbf{J}_\mathrm{S}\!+\!\mathbf{J}_\mathrm{M},
	\label{FIM}
\end{equation}
where $\mathbf{J}_\mathrm{S}$ and $\mathbf{J}_\mathrm{M}$ are the prior information FIM and the measurement FIM, respectively.
According to the state evolution model and measurement model, the FIMs can be readily computed as 
\begin{gather}
	\mathbf{J}_\mathrm{S}(\mathbf{x}_n)=( \mathbf{F}\mathbf{J}^{-1}_\mathrm{S}(\mathbf{x}_{n-1})\mathbf{F}^T + \mathbf{W} )^{-1}, \\
	\mathbf{J}_\mathrm{M} = \mathbf{H}_n^T \mathbf{R}_n^{-1} \mathbf{H}_n,
\end{gather}
where $\mathbf{H}_n = \left. \frac{\partial \mathbf{h}}{\partial \mathbf{x}} \right|_{\mathbf{x}=\mathbf{x}_n} $ is the jacobian matrix of $\mathbf{h}(\mathbf{x})$, which can be obtained as 
$\frac{\partial \mathbf{h}}{\partial \mathbf{x}} = \left[\frac{\partial d_1}{\partial \mathbf{x}},...,\frac{\partial d_M}{\partial \mathbf{x}},\frac{\partial v_1}{\partial \mathbf{x}},...,\frac{\partial v_M}{\partial \mathbf{x}} \right]^T$.

It can be found that the PCRB of the target state is the function with respect to the locations and bandwidth of UAVs. 
In addition, since the real target state is always unknown, one can only resort to predicted parameters for PCRB.
Therefore, given the predictions $\hat{\mathbf{x}}_{n|n-1} = \mathbf{F}\hat{\mathbf{x}}_{n-1}$ for the target state in the time interval $n$, the predicted PCRB of the target state can be written as
$\mathrm{tr}(\mathbf{C}(\hat{\mathbf{x}}_{n|n-1}) = f\left(\{\mathbf{q}_{m,n}\},\{\eta_{m,n}\}\right).$

\subsection{Dynamic Target Sensing Mechanism}
To obtain the latest measurements, the UAV trajectories and bandwidth allocation are optimized by minimizing the predicted PCRB.
Therefore, a dynamic target sensing mechanism is designed. 
To tackle the nonlinearity of the measurement model, the extended Kalman filtering (EKF) is used to update the target state
\begin{equation}
    \hat{\mathbf{x}}_n = \hat{\mathbf{x}}_{n|n-1} +\mathbf{K}_n(\mathbf{y}_n-\mathbf{h}(\hat{\mathbf{x}}_{n|n-1})),
\end{equation}
where $\mathbf{K}_n$ is the kalman gain.

With the prediction and update of the target state by EKF, the closed loop of the dynamic target sensing mechanism is formed.
Specifically, the trajectories and bandwidth allocation of UAVs can be adjusted dynamically through the repetition of 1) target state prediction, 2) UAVs trajectories and bandwidth allocation optimization, 3) UAVs performing sensing and communication, and 4) target state update.

\subsection{Problem Formulation}
Based on the previous discussions, we formulate the problem with the aim of minimizing the PCRB for the target state by optimizing the UAV trajectory, bandwidth allocation, and UAV-BS association, while satisfying the communication requirement.
Accordingly, the optimization problem in the $n$-th time interval is formulated as
\begin{align}
	\text{(P1):}& \min_{\mathbf{Q}_{n},\boldsymbol{\eta}_{n},\mathbf{A}_{n}} \mathrm{tr}(\mathbf{C}(\hat{\mathbf{x}}_{n|n-1})=f\left( \mathbf{Q}_{n},\boldsymbol{\eta}_{n}\right) \\ 
	\text{s.t.\ } 
	& R_{m,n}\left(\mathbf{q}_{m,n},\eta_{m,n},a_{m,k,n} \right) \geq R_{\mathrm{th}}, \forall m, \tag{14a} \label{26a} \\
	& \min_{k} \| \mathbf{q}_{m,n}-\mathbf{w}_k \| \leq d_\text{th}, \forall m,  \tag{14b} \label{26b}\\
	& \sum\nolimits_{m=1}^M \eta_{m,n} = 1,\tag{14c} \label{26c} \\
	& \eta_{m,n} \in [0,1], \forall m, \tag{14d} \label{26d} \\
	& \sum\nolimits_{m=1}^M a_{m,k,n} \leq N_a, \forall k,  \tag{14e} \label{26e}\\
	& \sum\nolimits_{k=1}^K a_{m,k,n} \leq 1, \forall m,  \tag{14f} \label{26f}\\
	& a_{m,k,n} \in \{0,1\}, \forall m,k, \tag{14g} \label{26g}\\
	& \Vert \mathbf{q}_{m,n}-\mathbf{q}_{m,n-1} \Vert \leq V_{\max}\Delta t, \forall m,  \tag{14h} \label{26h}\\
	& \Vert \mathbf{q}_{i,n}-\mathbf{q}_{j,n} \Vert \geq D_\text{s}, \forall i,j \in \mathcal{M}, i\neq j, \tag{14i} \label{26i}\\
	& \| \mathbf{q}_{m,n}-\hat{\mathbf{u}}^t_{n|n-1} \| \geq D_\text{s}, \forall m, \tag{14j} \label{26j}
\end{align}
Where $\mathbf{Q}_{n}=\{\mathbf{q}_{m,n}\}_{m=1}^M, \boldsymbol{\eta}_{n}=\{\eta_{m,n}\}_{m=1}^M$, and $\mathbf{A}_{n}=[a_{m,k,n}]_{M \times K}$.
$\hat{\mathbf{u}}^t_{n|n-1}=(\hat{x}^t_{n|n-1},\hat{y}^t_{n|n-1})$ is the predicted location of the target in the time interval $n$, $V_\mathrm{max}$ is the UAV maximum speed, $D_s$ is the safe distance between the UAVs for collision avoidance.
The communication requirement is constrained in (\ref{26a}), where $R_\mathrm{th}$ is the communication rate threshold of UAVs.
The constraint in (\ref{26b}) ensures the real-time and reliable control of the BS to the UAV, where $d_\text{th} \triangleq \sqrt{\frac{P^\mathrm{BS}_t \beta_0}{\sigma_0^2 \Gamma_\mathrm{th}}-\Delta H^2}$.
The maximum distance between two consecutive locations is constrained in (\ref{26h}).
The minimum safe distance for avoiding collision is limited in (\ref{26i}) and (\ref{26j}).

\section{Communication-Constrained Sensing Error Minimization Via Multi-UAV Collaboration}
In this section, we first obtain a communication-optimal solution to check the feasibility of problem (P1).
Then, to minimize PCRB, we employ an AO framework, where the UAV trajectory and bandwidth allocation are optimized by the descent direction search algorithm alternatively.

\subsection{Feasibility of Problem (P1)}
Before solving (P1), we first check its feasibility.
Note that (P1) is feasible if and only if the constraint $\text{(\ref{26a})}$ is satisfied, which implies that the maximum of $R_c=\min_{m} R_{m,n}$ is no smaller than the threshold $R_{\mathrm{th}}$.
Specifically, $R_c^{\max}$ can be obtained by solving the following optimization problem:
\begin{align}
	\text{(P1-F):}& \max_{\mathbf{Q}_{n},\boldsymbol{\eta}_{n},\mathbf{A}_{n}} R_c \\ 
	\text{s.t.\ } 
	& (\text{\ref{26b}})-(\text{\ref{26j}}). \notag
\end{align}

\textbf{Lemma 1}: 
Given ${\{\mathbf{q}_{m,n}\}}$, ${\{a_{m,k,n}\}}$, the communication-optimal bandwidth allocation can be obtained as
\begin{equation}
	\eta_{m,n}^{\text{c}*} = \frac{1}{\widetilde{R}_{m,n} \sum_{m=1}^M\frac{1}{\widetilde{R}_{m,n}}}, m\in[1,M],
\end{equation}
where $\widetilde{R}_{m,n} = B\log_2 \left(1+\frac{\rho_0}{\|\mathbf{q}_{m,n}-\mathbf{w}_{k_{m,n}} \|^2+ \Delta H^2}\right)$, 
$\rho_0=\frac{P^\text{u}_t \beta_0}{\sigma_0^2}$,
and $k_{m,n} = \{k|a_{m,k,n}=1,k\in[1,K]\}$ is the associated BS of UAV $m$.

\emph{Proof}: 
Define $R_c = \min_m R_{m,n}$, the maximum value of $R_c$ can be achieved at $R_{m,n}=R_c, \forall m$, i.e., $\eta_{m,n}^{\text{c}*}\widetilde{R}_{m,n}=R_c, \forall m$, and $\eta_{m,n}^{\text{c}*} = \frac{R_c}{\widetilde{R}_{m,n}}$.
Since $\sum_{m=1}^M \eta_m=1$, thus $\sum_{m=1}^M \frac{R_c}{\widetilde{R}_{m,n}}=1$, then we have $\eta_{m,n}^{\text{c}*} = \frac{1}{\widetilde{R}_{m,n} \sum_{m=1}^M\frac{1}{\widetilde{R}_{m,n}}}$.
\hfill $\blacksquare$

According to Lemma 1, the objective function of (P1-F) can be rewritten as $R_c^* = 1/{\sum_{m=1}^M\frac{1}{\widetilde{R}_{m,n}}}$.
It is noted that maximizing $R_c^*$ is equivalent to minimize $\sum_{m=1}^M\frac{1}{\widetilde{R}_{m,n}}$.

\textbf{Lemma 2}: 
The communication-optimal UAV-BS association is $a_{m,k,n}^{\text{c}*} = a_{m,k,n-1}^{\text{c}*}$.

\emph{Proof}: 
At the $(n-1)$-th time interval, for communication-optimal UAV-BS association $a_{m,k,n-1}^{\text{c}*}$, we have $\| \mathbf{q}_{m,n-1}- \mathbf{w}_{k^{\text{c}*}_{m,n-1}}\| \leq \|\mathbf{q}_{m,n-1}- \mathbf{w}_{\tilde{k}_{m,n-1}}\|$, where $\tilde{k}_{m,n-1}$ is the other BSs that UAV $m$ can access.
Thus, the UAV flying towards the communication-optimal associated BS at the previous time interval can achieve the maximum communication gain, which completes the proof.
\hfill $\blacksquare$

With the known $\{\mathbf{q}_{m,n-1}\}$, we can calculate $b_{m,k,n-1} = 1/\left({B\log_2\left( 1+\frac{\rho_0}{\| \mathbf{q}_{m,n-1}-\mathbf{w}_k\|^2 + \Delta H^2}\right)}\right), \forall m,k$.
Then the communication-optimal UAV-BS association can be obtained by solving the following problem:
\begin{align}
	\text{(P1-F.1):}& \min_{\{a_{m,k,n}\}} \sum_{m=1}^M \sum_{k=1}^K a_{m,k,n}b_{m,k,n-1} \\ 
	\text{s.t.\ } 
	& (\text{\ref{26e}})-(\text{\ref{26g}}). \notag
\end{align}

Since the constraints involve the binary variables, we relax $a_{m,k,n}$ to $0 \leq a_{m,k,n} \leq 1$.
Thus, the problem (P1-F.1) can be transformed into a standard linear program, and the continuous solution can be obtained by using the CVX solver.
To construct the final binary solution, we set the element in $\{a_{m,k,n}\}_{k=1}^{K}$ with the maximum value to 1 and the other elements to 0.

For the given UAV-BS association $\{a_{m,k,n}\}$, to achieve the maximum communication rate, the UAV will fly towards the associated BS at the maximum speed. 
Thus the communication-optimal UAV trajectory can be expressed as:
\begin{equation}
	\mathbf{q}_{m,n}^{c*}\!=\!
	\begin{cases}
		\! \mathbf{w}_{k_{m,n}},\left\|\mathbf{w}_{k_{m,n}}-\mathbf{q}_{m,n-1}\right\|\leq V_{\max}\Delta t, \\
		\! \mathbf{q}_{m,n-1}\!+\!V_{\max}\Delta t \! \frac{\mathbf{w}_{k_{m,n}}\!-\!\mathbf{q}_{m,n-1}}{\left\|\mathbf{w}_{k_{m,n}}\!-\!\mathbf{q}_{m,n-1}\right\|}\!,\! otherwise.
	\end{cases}
\end{equation}

Based on the obtained communication-optimal solution, the communication rate of UAV $m$ can be obtained as $R_{m,n}\left(\mathbf{q}_{m,n}^{c*},\eta_{m,n}^{c*},a_{m,k,n}^{c*} \right)=R_c^* $. Therefore, the problem (P1) is feasible only if $R_c^* \geq R_\text{th}$.

\subsection{Proposed Algorithm for ISAC Solution}
Since the PCRB objective function is related to the UAV trajectory and bandwidth allocation, to obtain a larger solution space of $\mathbf{Q}_{n}$ and $\boldsymbol{\eta}_n$, the UAV-BS association in (P1) is given by the $a_{m,k,n}^{c*}$ obtained in Section III-A.
Then, we adopt AO framework to facilitate problem solving, the minimization process of PCRB consists of the alternating iterative optimization of $\mathbf{Q}_{n}$ and $\boldsymbol{\eta}_n$.
We first divide the problem (P1) into two subproblems, which are given by
\begin{align}
	\text{(P1.1):}& \min_{\mathbf{Q}_{n}} f\left(\mathbf{Q}_{n}\right) \\ 
	\text{s.t.\ } 
	& \| \mathbf{q}_{m,n}-\mathbf{w}_{k_{m,n}} \|^2 \leq \frac{\rho_0}{2^{\frac{R_{\mathrm{th}}}{\eta_{m,n}B}}-1}-\Delta H^2, \forall m, \tag{19a} \label{45a} \\
	& (\text{\ref{26b}}), (\text{\ref{26h}})-(\text{\ref{26j}}), \notag
\end{align}
\begin{align}
	\text{(P1.2):}& \min_{\boldsymbol{\eta}_n} f\left(\boldsymbol{\eta}_n\right) \\ 
	\text{s.t.\ } 
	& \eta_{m,n} \!\geq\! \frac{R_{\mathrm{th}}}{B\log_2\left(1\!+\!\frac{\rho_0}{\| \mathbf{q}_{m,n}-\mathbf{w}_{k_{m,n}} \|^2+\Delta H^2}\right)}, \forall m, \tag{20a} \label{46a}\\
	& (\text{\ref{26c}}), (\text{\ref{26d}}), \notag
\end{align}
where the constraints in (\text{\ref{45a}}) and (\text{\ref{46a}}) are transformed from the constraint in (\text{\ref{26a}}).

Since $f\left(\mathbf{Q}_{n}\right)$ and $f\left(\boldsymbol{\eta}_n\right)$ are non-convex and difficult to express in closed form, a descent direction search method is proposed to obtain the solution.

Firstly, for (P1.1), we approximate $f\left(\mathbf{Q}_{n}\right)$
by its first-order Taylor expansion at $\mathbf{Q}^{(l)}_{n}$ to obtain its descent direction as 
\begin{equation}
	\begin{aligned}
		f\left(\mathbf{Q}_n\right)\! = 
		&f(\mathbf{Q}^{(l)}_{n})\! +\! \sum_{m=1}^M \nabla f_{x^u_{m,n}}(\mathbf{Q}^{(l)}_{n})(x^u_{m,n}\!-\!{x}^{u{(l)}}_{m,n}) \\
		& +\sum_{m=1}^M \nabla f_{y^u_{m,n}}(\mathbf{Q}^{(l)}_{n})(y^u_{m,n}-y^{u{(l)}}_{m,n}),
	\end{aligned}
	\label{f1}
\end{equation}
where $\mathbf{Q}_n^{(l)}$ is $\mathbf{Q}_n$ obtained from the $l$-th iteration, 
$\nabla f_{x^u_{m,n}}(\cdot)$ and $\nabla f_{y^u_{m,n}}(\cdot)$ denote the gradient of $f(\cdot)$ with respect to $x^u_{m,n}$ and $y^u_{m,n}$.
By minimizing the right-hand-side (RHS) of (\text{\ref{f1}}), we can find the descent direction of $f\left(\mathbf{Q}_{n}\right)$. 
Since the first term is fixed, the minimizing process focuses on the last two terms of the RHS.
By updating the RHS iteratively from the obtained descent direction, the local optimal solution of (P1.1) can be gradually approached.
The iterative form of (P1.1) can be reformulated as
\begin{align}
	\text{(P1.1.1):}& \min_{\mathbf{Q}_{n}} g(\mathbf{Q}_n)  \\ 
	\text{s.t.\ } 
	& (\text{\ref{45a}}), (\text{\ref{26b}}), (\text{\ref{26h}})-(\text{\ref{26j}}), \notag
\end{align}
where $g(\mathbf{Q}_n) = \sum_{m=1}^M \nabla f_{x_{m,n}}(\mathbf{Q}^{(l)}_{n})(x^u_{m,n}-x^{u{(l)}}_{m,n}) + \sum_{m=1}^M \nabla f_{y_{m,n}}(\mathbf{Q}^{(l)}_{n})(y^u_{m,n}-y^{u{(l)}}_{m,n})$.
By solving (P1.1.1), we can obtain the optimal solution of $\mathbf{Q}_n$, denoted as $\mathbf{Q}^*_n$.
Thus the descent direction of $f(\mathbf{Q}_n)$ can be calculated as $\mathbf{Q}^*_n-\mathbf{Q}_n^{(l)}$.
Then we search the solutions along $\mathbf{Q}^*_n-\mathbf{Q}_n^(l)$ with a stepsize $\omega(\omega \in[0,1])$ and find $\omega^*$ that minimizes $f(\mathbf{Q}_n)$.
Thus the next expansion point of $f(\mathbf{Q}_n)$ is
\begin{equation}
	\mathbf{Q}^{(l+1)}_n=\omega^*(\mathbf{Q}^*_n-\mathbf{Q}^{(l)}_n)+\mathbf{Q}^{(l)}_n.
\end{equation}
Then the $\mathbf{Q}^{(l+1)}_n$ is used to the next iteration.

However, (P1.1.1) cannot be solved directly since the nonconvex constraints in (\text{\ref{26b}}), (\text{\ref{26i}}) and (\text{\ref{26j}}).
For the constraint (\text{\ref{26b}}), define $k'_{m,n-1} \triangleq \arg\min_k \|\mathbf{q}_{m,n-1} - \mathbf{w}_k \|$, which denotes the BS closest to the $m$-th UAV at the time interval $n-1$.
It can be found that the constraint (\text{\ref{26b}}) actually denotes the feasible region of $\mathbf{q}_{m,n}$, i.e., the union of the circular areas with a radius of $d_\text{th}$ centered on BSs.
Combined with the constraint (\text{\ref{26h}}), the feasible region of $\mathbf{q}_{m,n}$ can be further contracted.
Then, for the case 1 that $\|\mathbf{q}_{m,n-1}-\mathbf{w}_{k'_{m,n-1}}\| \leq d_\text{th}-V_{\max}\Delta t$, $\mathbf{q}_{m,n}$ must satisfy the constraint (\text{\ref{26b}}) since $\|\mathbf{q}_{m,n-1}-\mathbf{w}_{k'_{m,n-1}}\| \leq d_\text{th}$.
For the case 2 that $d_\text{th}-V_{\max}\Delta t <\|\mathbf{q}_{m,n-1}-\mathbf{w}_{k'_{m,n-1}}\| \leq d_\text{th}$, the constraint (\text{\ref{26b}}) can be approximated as $\|\mathbf{q}_{m,n}-\mathbf{w}_{k'_{m,n-1}}\| \leq d_\text{th}$ since it dominates the feasible region.
Therefore, the constraint (\text{\ref{26b}}) can be approximated as 
\begin{equation}
	\|\mathbf{q}_{m,n}-\mathbf{w}_{k'_{m,n-1}}\| \leq d_\text{th}, m\in[1,M].
	\label{d_ba}
\end{equation}

By applying the first-order Taylor expansion in the $r$-th iteration, the constraints (\text{\ref{26i}}) and (\text{\ref{26j}}) can be rewritten as 
\begin{equation}
	2\left(\mathbf{q}^{(r)}_{i,n}\!-\!\mathbf{q}^{(r)}_{j,n}\right)^T \left(\mathbf{q}_{i,n}-\mathbf{q}_{j,n}\right)-\Vert \mathbf{q}^{(r)}_{i,n}-\mathbf{q}^{(r)}_{j,n} \Vert^2 \geq D_\text{s}^2,
	\label{safe1}
\end{equation}
\begin{equation}
	\| \mathbf{q}^{(r)}_{m,n}\!-\!\hat{\mathbf{u}}^s_{n|n-1} \|^2 \!+\!2\left(\mathbf{q}^{(r)}_{m,n}\!-\!\hat{\mathbf{u}}^s_{n|n-1}\right)^T \! \left(\mathbf{q}_{m,n} \!-\! \mathbf{q}^{(r)}_{m,n}\right)\! \geq \! D_\text{s}^2.
	\label{safe2}
\end{equation}

With the above manipulations, the upper-bounded solution to (P1.1.1) can be obtained by addressing the convex problem
\begin{align}
	\text{(P1.1.2):}& \min_{\mathbf{Q}_{n}} g(\mathbf{Q}_n) \\ 
	\text{s.t.\ } 
	& (\text{\ref{45a}}), (\text{\ref{d_ba}}), (\text{\ref{26h}}), (\text{\ref{safe1}}), (\text{\ref{safe2}}), \notag
\end{align}
Then the solution to (P1.1.1) can be obtained by solving (P1.1.2) iteratively.
Therefore, a solution that minimizes the PCRB can be obtained. The descent direction search algorithm for (P1.1) is shown in \textbf{Algorithm 1}. 

\begin{algorithm}[t]
	\footnotesize
	\caption{ \small The Descent Direction Search Algorithm for (P1.1)}
	\begin{algorithmic}[1]
		\STATE Initialize $\mathbf{Q}^{(0)}_{n} = \mathbf{Q}_{n-1}$, $l=0$;
		\REPEAT
		\STATE Formulate (P1.1.2), solve (P1.1.1) by using SCA, and obtain $\mathbf{Q}^*_n$;
		\FOR{$\omega=0:\Delta_{\omega}:1$}
		\STATE Calculate $f(\mathbf{Q}'_n)$, where $\mathbf{Q}'_n=\omega(\mathbf{Q}^*_n-\mathbf{Q}^{(l)}_n)+\mathbf{Q}^{(l)}_n$;
		\STATE Find $\omega$ that minimizes $f(\mathbf{Q}_n)$ and denote it as $\omega^*$;
		\ENDFOR
		\STATE Obtain $\mathbf{Q}^{(l+1)}_n=\omega^*(\mathbf{Q}^*_n-\mathbf{Q}^{(l)}_n)+\mathbf{Q}^{(l)}_n$;
		\STATE Update $l=l+1$;
		\UNTIL{$\frac{f(\mathbf{Q}^{(l)}_n) - f(\mathbf{Q}^{(l+1)}_n)}{f(\mathbf{Q}^{(l)}_n)} \leq \varepsilon$.}
	\end{algorithmic}
\end{algorithm}

Similarly, the same solving process is applied to the problem (P1.2), the iterative form of (P1.2) can be formulated as 
\begin{align}
	\text{(P1.2.1):}& \min_{\boldsymbol{\eta}_n} \sum_{m=1}^M \nabla f_{\eta_{m,n}}\left(\boldsymbol{\eta}^l_n\right)(\eta_{m,n}-\eta^l_{m,n}) \\ 
	\text{s.t.\ } 
	& (\text{\ref{46a}}),(\text{\ref{26c}}),(\text{\ref{26d}}). \notag
\end{align}
(P1.2.1) is a convex problem and can be solved directly.
The detailed procedure to search for a solution is similar to Algorithm 1.

Finally, based on the AO framework, the solution to (P1) can be obtained by alternately optimizing the UAV trajectory and bandwidth allocation.
The iterative algorithm for (P1.1) is described in \textbf{Algorithm 2}.

\emph{Computational complexity analysis}: 
The overall computational complexity of Algorithm 2 is dominated by the AO loop and the inner subproblem solvers. In each AO iteration, the trajectory optimization subproblem (solved by Algorithm 1) involves solving a convex problem with 
$\mathcal{O}(M)$ variables, yielding a per-iteration complexity of $\mathcal{O}(M^3)$. The bandwidth allocation subproblem is also convex with $\mathcal{O}(M)$ variables, solvable in $\mathcal{O}(M^3)$ time. Assuming $I_\text{AO}$ AO iterations and $I_\text{Alg1}$ inner iterations for Algorithm 1, the total complexity of Algorithm 2 is approximately $\mathcal{O}(I_\text{AO} \cdot I_\text{Alg1} \cdot M^3)$. This polynomial complexity is relatively low and feasible for real-time implementation with a moderate number of UAVs. 

\begin{algorithm}[t]
	\footnotesize
	\caption{\small Proposed Iterative Algorithm for (P1)}
	\begin{algorithmic}[1]
		\STATE Solve (P1-F.1), obtain the UAV-BS association $\mathbf{A}_n$.
		\STATE Initialize $\mathbf{Q}^{(0)}_{n} = \mathbf{Q}_{n-1}$, $\boldsymbol{\eta}^{(0)}_n = \boldsymbol{\eta}_{n-1}$, $l=0$.
		\REPEAT
		\STATE Solve (P1.1) by algorithm 1, obtain the optimized UAV trajectory $\mathbf{Q}^{(l+1)}_n$.
		\STATE Solve (P1.2), obtain the optimized bandwidth allocation $\boldsymbol{\eta}^{(l+1)}_n$.
		\STATE Caculate the objective value of (P1), $f^{(l+1)} = f(\mathbf{Q}^{(l+1)}_n, \boldsymbol{\eta}^{(l+1)}_n)$.
		\STATE Update $l=l+1$.
		\UNTIL{$\frac{f^{(l)} - f^{(l+1)}}{f^{(l)} } \leq \varepsilon$.}
	\end{algorithmic}
\end{algorithm}

\section{Simulation and Results Analysis}
In this section, numerical results are provided to validate the proposed algorithm. We consider a geographical region of size $3\  \text{km} \times 3 \ \text{km}$, where $K=5$ BSs are deployed randomly and $M=3$ UAVs perform ISAC tasks.
Unless otherwise stated, the maximum UAV number that one BS can associate simultaneously is set as $N_a$= 3, the communication requirement of UAV is set as $R_\text{th}$ = 10 Mbps, UAV maximum speed is set as $V_\text{max}$ = 25 m/s, $\kappa$ = 0.5, $\alpha_d=3.6\times 10^{-8}$, $\alpha_v= 1.4\times 10^{-9}$.
We set the initial state of the target as $\mathbf{x}_0=[600,1600,30,0]^T$, the initial locations of UAVs are set as $(1500, 2300)$, $(807, 1100)$, and $(2193, 1100)$.
The other parameters are shown in Table \ref{paras}, and the simulation tool is MATLAB. 

\begin{table}[t]
	\scriptsize
	\centering
	\caption{Parameter Settings}
	\begin{tabular}{|c|c|c|c|c|c|} \hline
		Parameter & Value  & Parameter & Value & Parameter & Value\\ \hline \hline
		$P_t^\text{u}$ & 0.1 W  & $P_t^\text{BS}$ & 1 W & $H$ & 100 m \\ \hline
		$B$ & 20 MHz & $f_c$ & 28 GHz & $H^\text{B}$ & 15 m \\ \hline
		$\sigma_0^2$ & -130 dBm & $\sigma_{RCS}$ & -10 dBsm & $T$ & 60 s \\ \hline
		$D_\text{s}$ & 30 m & $\Gamma_{\text{th}}$ & 11 dB & $\Delta t$ & 1 s \\ \hline
	\end{tabular}
	\label{paras}
\end{table}

\begin{figure}[t]
	\centering
	\subfloat[UAVs' trajectories.]{\includegraphics[width=5.5cm]{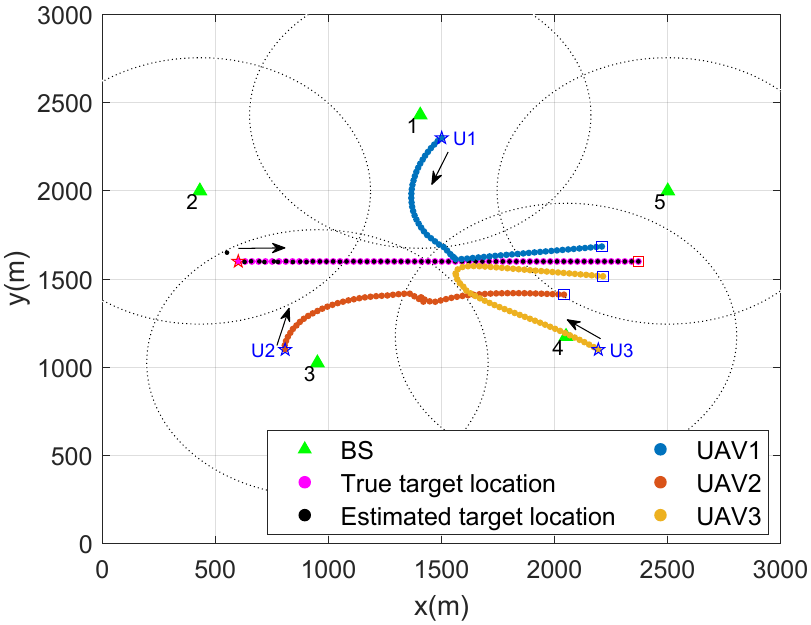}} \\
	\subfloat[Bandwidth allocation.]{\includegraphics[width=5cm]{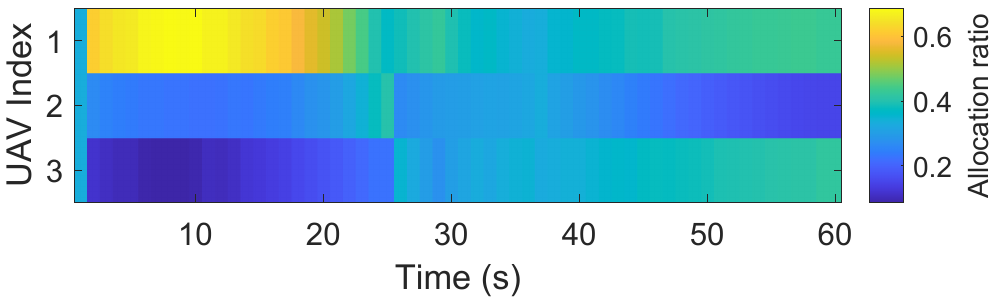}}
	\caption{Optimized UAVs' trajectories and bandwidth allocation}
    \vspace{-0.5cm}
	\label{trajectory and bandwidth}
\end{figure}

Fig. \ref{trajectory and bandwidth} shows the optimized UAVs' trajectories and bandwidth allocation obtained from the proposed algorithm.
In Fig. \ref{trajectory and bandwidth}(a), it can be seen that the UAVs all approach the target at a fast speed, the main reason is that the short distance between UAV and the target leads to a higher sensing SNR, then lower sensing error can be obtained.
When the UAV is relatively close to the target, it no longer flies in the direction closest to the target but gradually increases the azimuth angle between the UAV and the target. The main reason is that the positioning of the target also depends on the geometric position distribution of the UAV relative to the target.
In Fig. \ref{trajectory and bandwidth}(b), it can be found that the bandwidth resources tend to be allocated to UAVs with better geometric positions, since these UAVs can achieve greater sensing performance gains with limited resources.

For evaluation, the proposed is compared with three benchmarks: Static Deployment (SD), Average Bandwidth Allocation (ABA), and CRB-criterion based algorithm.
In SD, UAVs are deployed statically in the initial locations and the bandwidth allocation is optimized.
ABA optimized the trajectories of UAVs by the proposed algorithm 1.
The CRB-criterion based algorithm uses CRB as the objective function for optimization.

\begin{figure}[t]
	\centering
	\subfloat[PCRB.]{\includegraphics[width=4.5cm]{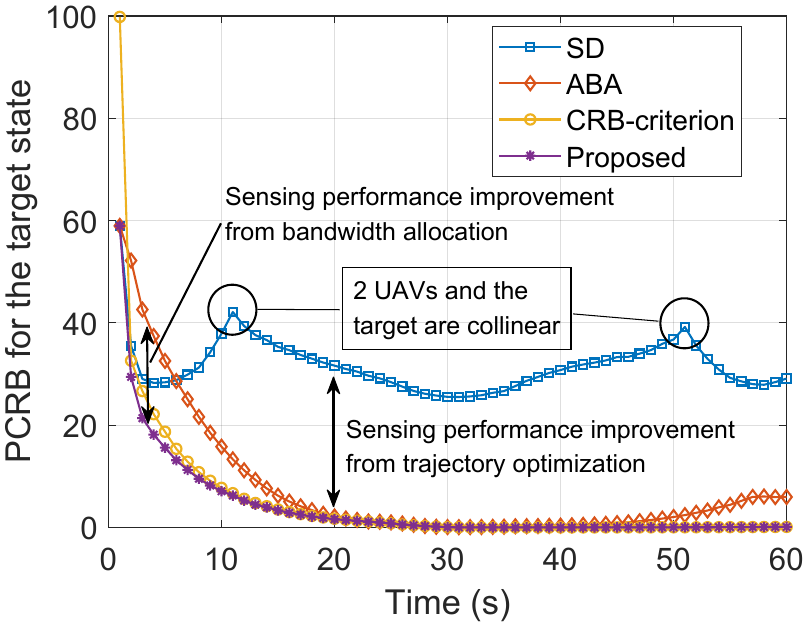}} 
	\subfloat[Target state error.]{\includegraphics[width=4.6cm]{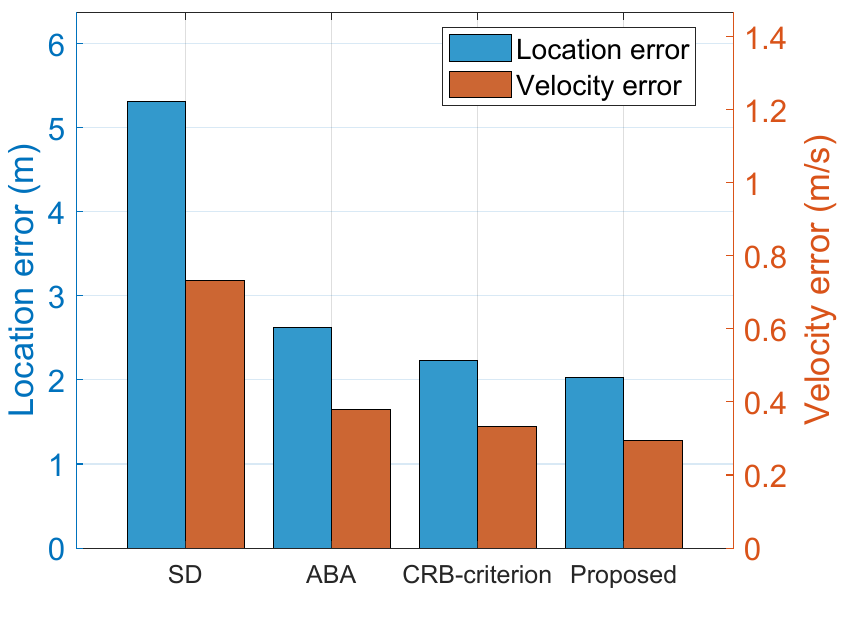}}
	\caption{Sensing performance comparison}
    \vspace{-0.3cm}
	\label{sensing performance}
\end{figure}

Fig. \ref{sensing performance} shows the sensing performance comparison among different algorithms.
It can be observed from Fig. \ref{sensing performance}(a) that the proposed algorithm can achieve the minimum PCRB in the whole mission.
Compared with SD and ABA, the joint optimization of UAVs' trajectories and bandwidth allocation can improve sensing performance effectively.
Since the CRB-criterion method only changes the objective function, it exhibits similar performance to the proposed algorithm.
To describe the sensing error intuitively, Fig. \ref{sensing performance}(b) shows the root mean squared error (RMSE) of target's location and velocity obtained from different algorithms.
It can be found that the proposed algorithm can obtain the minimum sensing error.
Compared with different benchmarks, the location error of the proposed algorithm is 61.1\%, 22.8\% and 9.0\% lower than that of the benchmarks, respectively.
The velocity error is reduced by 59.9\%, 29.1\% and 12.0\%, respectively.

\vspace{-0.1cm}
\section{Conclusion}
\vspace{-0.1cm}
In this paper, an ISAC-enabled multi-UAV dynamic collaborative target sensing scheme was proposed to support the development of LAE.
Specifically, multiple UAVs can exploit ISAC signals to collaboratively conduct sensing of the mobile target while communicating with the terrestrial BSs.
To achieve the precise sensing of the dynamic target, the PCRB for the target state was derived as the sensing error metric.
Then, we formulated an optimization problem aiming at minimizing the PCRB by optimizing the UAV-BS association, UAVs' trajectories and bandwidth allocation.
To tackle the formulated non-convex problem, we proposed a low-complexity iterative algorithm to solve it efficiently. 
First, a communication-optimal solution was obtained to check the feasibility and determine the UAV-BS association.
Then, AO framework is employed to alternatively optimize the UAVs’ trajectories and bandwidth allocation based on the descent direction search algorithm.
The simulation results show the effectiveness and superiority of our proposed designs as compared to various benchmarks.
In the future, we look forward to exploring the use of the ISAC multi-UAV collaborative network for multi-target tracking and integrating computing capabilities to provide more robust and intelligent services.

\vspace{-0.25cm}
\bibliographystyle{IEEEtran}
\bibliography{IEEEabrv,reference}

\end{document}